\documentclass[aps,prc,showpacs,nofootinbib]{revtex4}

\usepackage{epsfig}
\usepackage{graphicx}

\begin{document}

\title{Complete experiment in $e^{\mp}N$ elastic scattering, in presence of two-photon exchange}

\author{M. P. Rekalo}
\altaffiliation{Permanent address:
\it NSC Kharkov Physical Technical Institute, 61108 Kharkov, Ukraine}
\affiliation{\it DAPNIA/SPhN, CEA/Saclay, 91191 Gif-sur-Yvette Cedex,
France }

\author{E. Tomasi-Gustafsson}
\affiliation{\it DAPNIA/SPhN, CEA/Saclay, 91191 Gif-sur-Yvette Cedex,
France }

\date{\today}

\begin{abstract}
We propose  possible methods to measure the nucleon electromagnetic form factors  in presence of two-photon exchange. Using a general parametrization of the spin structure of the matrix element for elastic $eN$-scattering, in terms of three independent complex amplitudes, we show that the  measurements of the differential cross section for electron and positron nucleon elastic scattering, in the same kinematical conditions, allows to extract the nucleon electromagnetic form factors. The same is correct for the polarization method, with the measurement of the $P_{x,z}$ components of the final nucleon polarization (for the scattering of longitudinally polarized electrons and positrons). An alternative way, in absence of positron beam, is to measure a definite set of T-odd  polarization observables, which includes three different experiments or only T-even observables, with five independent experiments. In both cases, the ratio $G_E(Q^2)/G_M(Q^2)$ is related to quantities of the order of $\alpha$, which requires different polarization experiments of very high accuracy.
\end{abstract}

\maketitle
\section{Introduction}

Electron hadron scattering is considered to be the best way to study the hadron structure. The underlying mechanism in elastic (and inelastic) scattering is the  one-photon exchange which can be exactly calculated in quantum electrodynamics (QED), at least for the lepton vertex.  In this respect the electromagnetic probes are traditionally preferred to the hadronic beams for the investigation of nucleons and light nuclei \cite{Ho62}. Radiative corrections, firstly calculated by Schwinger \cite{Shwinger} are important for the discussion of the experimental determination of the differential cross section and are also calculable in QED. 

However, due to the steep decreasing of the electromagnetic form factors with $Q^2$, the two-photon exchange mechanism where the momentum transfer, $Q^2$, is shared between the two photons, can become important with increasing $Q^2$. This fact was already indicated in the seventies \cite{Gu73}, but it was never experimentally observed.

Numerous tests of the validity of the one-photon mechanism have been done in the past, using different methods: test of the linearity of the Rosenbluth formula for the differential cross section, comparison of the $e^+p$ and $e^-p$-cross sections, attempts to measure various T-odd polarization observables, but no effect was visible beyond the precision of the experimental data.

Note that the two-photon exchange should appear at smaller $Q^2$ for heavier targets: $d$, $^3\!He$,  $^4\!He$, because the corresponding form factors decrease faster with $Q^2$ in comparison with protons. In \cite{Re99} the possible effects of $2\gamma$-exchange have been estimated from  the precise data on the structure function $A(Q^2)$, obtained at JLab in electron deuteron elastic scattering, up to $Q^2=6$ GeV$^2$ \cite{Al99,Ab99}. The possibility of $2\gamma$-corrections has not been excluded by this analysis, starting from $Q^2=1$ GeV$^2$, and the necessity of  dedicated experiments was pointed out. From this kind of consideration, one would expect to observe the two-photon contribution in $eN$-scattering at larger momentum transfer, for $Q^2\simeq 10$ GeV$^2$.

In a previous paper \cite{Re03}, on the line of the $ed$-analysis \cite{Re99}, we proved that general properties of the hadron electromagnetic interaction, such as the C-invariance and the crossing symmetry, give rigorous prescriptions for different observables for the elastic scattering of electrons and positrons by nucleons. These concrete prescriptions help in identifying a possible manifestation of the two-photon exchange mechanism. 

Recent developments in the field of hadron electromagnetic form factors (FFs) are due to the very precise and surprising data obtained at the Jefferson Laboratory (JLab), in $\vec e+p\to e+\vec p$ elastic scattering \cite{Jo00}, based on the polarization transfer method \cite{Re68} which show that the electric and magnetic distributions in the proton are different.

The existing data show a discrepancy between the $Q^2$-dependence of the ratio
$R= \mu_p G_{Ep}$/$G_{Mp}$ of the electric to the magnetic proton form factors ($\mu_p$=2.79 is the proton magnetic moment), whether derived with the standard Rosenbluth separation \cite{Ar75} or with the polarization method. 

It has been suggested \cite{Br03} that, under specific assumptions, the presence of the 2$\gamma$-mechanism can explain such discrepancy. Therefore a careful experimental and theoretical analysis of this problem is necessary. The important point here is the calculation of radiative corrections to the differential cross section and to polarization observables in elastic $eN$-scattering. If these corrections are large (in absolute value) for the differential cross section \cite{Mo69}, in particular for high resolution experiments, a simplified estimation of radiative corrections to polarization phenomena \cite{Ma00} shows that radiative corrections are small for the ratio $P_L/P_T$ of longitudinal to transverse polarization of the proton emitted in the elastic collision of polarized electrons with an unpolarized proton target.

The purpose of this paper is to suggest a model independent strategy to extract electromagnetic nucleon form factors and to determine the two-photon contribution, on the basis of a general analysis of polarization phenomena in elastic $eN$-scattering. We discuss also the possibility to solve this problem, having a positron beam. We will show that the measurement of the differential cross section and of the components of the proton polarization in $e^-p$ and $e^+p$ elastic scattering (in identical kinematical conditions) is the most direct way to access the nucleon FFs and the $2\gamma$-contribution. In the present situation, when a  positron beam is absent, we show that an alternative way can be the measurements of a larger set of polarization observables in $e^-p$-scattering. Such experiments can be in principle actually performed at Jefferson Laboratory.

\section{Crossing symmetry and C-invariance}

The starting point of our analysis is the following general parametrization  of the spin structure of the matrix element for elastic $eN$-scattering, according to the formalism of \cite{Go57}:
\begin{equation}
{\cal  M}=\displaystyle\frac{e^2}{Q^2}\overline{u}(k_2)\gamma_{\mu}u(k_1)
\overline{u}(p_2) \left [{\cal  A}_1(s,Q^2)\gamma_{\mu}-{\cal  A}_2(s,Q^2) 
\displaystyle\frac{\sigma_{\mu\nu}q_{\nu}}{2m}+{\cal  A}_3(s,Q^2)\hat K {\cal  P}_{\mu}\right] u(p_1),
\label{eq:mat1}
\end{equation}
$$K=\displaystyle\frac{k_1+k_2}{2},~{\cal  P}=\displaystyle\frac{p_1+p_2}{2},$$
where ${\cal  A}_1-{\cal  A}_3$ are the corresponding invariant amplitudes, $k_1$ $(p_1)$ and $k_2$ $(p_2)$ are the four-momenta of the initial and final electron (nucleon), $m$ is the nucleon mass, $q=k_1-k_2$, $Q^2=-q^2>0$.
In case of one-photon exchange 
$$
{\cal  A}_1(s,Q^2)\to F_{1N}(Q^2),~
{\cal  A}_2(s,Q^2)\to F_{2N}(Q^2),~
{\cal  A}_3\to 0.
$$
$F_{1N}$ and $F_{2N}$ are the Dirac and Pauli nucleon electromagnetic form factors, which are real functions of the variable $Q^2$ - in the space-like region of momentum transfer. The same form factors describe also the one-photon mechanism for the elastic scattering of positrons by nucleons.

The spin structure of the matrix element, Eq. (\ref{eq:mat1}), has been established in analogy with elastic $np$-scattering \cite{Go57}, using the general properties of the electron-hadron interaction, such as the P-invariance and the relativistic invariance. Taking into account the identity of the initial and final states and the T-invariance of the electromagnetic interaction, the processes $e^{\mp}N\to e^{\mp} N$, in which  four  particles with spin 1/2 participate, are characterized by six independent products of four-spinors, describing the initial and final fermions.  The corresponding (model independent) parametrization of the matrix element can be done in many different but equivalent forms, in terms of six invariant complex amplitudes, ${\cal A}_i(s,Q^2)$, $i=1-6$, which are functions of two independent variables,  and  $s=(k+p_1)^2$ is the square of the total energy of the colliding particles. 

In principle, another set of variables can be considered: $\epsilon$ and $Q^2$, which is equivalent  to $s$ and $Q^2$ (in Lab system):
$$\epsilon^{-1}=1+2(1+\tau)\tan ^2\displaystyle\frac{\theta_e}{2},$$
where $\theta_e$ is the electron scattering angle in the laboratory (Lab) system, $\tau=Q^2/(4m^2)$. The variables $\epsilon$ and $Q^2$ are well adapted to the description of the properties of one-photon exchange for elastic $eN$-scattering, because, in this case, only the $Q^2$-dependence of the form factors has a dynamical origin, whereas the linear $\epsilon$-dependence of the differential cross section is a trivial consequence of the one-photon mechanism. On the other hand, the variables $s$ and $Q^2$ are
better suited to the analysis of the implications from crossing symmetry.

The conservation of the lepton helicity, which is a general property of leptons (at high energies), reduces the number of invariant amplitudes for elastic $eN$-scattering,  from six to three.

In the general case (with multi-photon exchanges) the situation is more complicated, because:
\begin{itemize}
\item The amplitudes ${\cal  A}_i(s,Q^2)$, $i=1-3$, are complex functions of two independent variables, $s$ and $Q^2$. 
\item The set of amplitudes ${\cal  A}_i^{(-)}(s,Q^2)$ for the process $e^-+N\to e^-+N$ is different from the set ${\cal  A}_i^{(+)}(s,Q^2)$ of corresponding amplitudes for positron scattering,  $e^++N\to e^++N$, which means that the properties of the scattering of positrons can not be derived from  ${\cal  A}_i^{(-)}(s,Q^2)$, as in case of the one-photon mechanism.
\item The connection of the amplitudes  ${\cal  A}_i(s,Q^2)$ with the nucleon electromagnetic form factors, $F_{iN}(Q^2)$, is non-trivial, because these amplitudes depend on many quantities, as, for example, the form factors of the $\Delta$-excitation - through the amplitudes of the virtual Compton scattering.
\end{itemize}

In this framework, the simple and transparent phenomenology of electron-hadron physics does not hold anymore, and in particular, it would  be very difficult  to extract information on the internal structure of a hadron in terms of electromagnetic form factors, which are real functions of one variable, from electron scattering experiments.

In the following text, we will show that the situation is not so involved, and that even in case of two-photon exchange, one can still use the formalism of form factors, if one takes into account the C-invariance of the electromagnetic interaction of hadrons.

In summary both reactions $e^{\mp}+N\to e^{\mp}+N$ are fully described by eight different real quantities:
\begin{itemize}
\item two real form factors $G_{E,M}(Q^2)$, which are functions of one variable, only,
\item three  functions: $\Delta G_E(Q^2,\epsilon)$,$\Delta G_M(Q^2,\epsilon)$,${\cal A}_3(Q^2,\epsilon)$, which are, generally, complex functions of two variables, $Q^2$ and $\epsilon$. 
\end{itemize}
We will use the following notations:
\begin{eqnarray*}
\widetilde{G}_E^{(\mp)}(Q^2,\epsilon)&=&{\cal  A}_1^{(\mp)}+\tau {\cal  A}_2^{(\mp)}(Q^2,\epsilon),\\
\widetilde{G}_M^{(\mp)}(Q^2,\epsilon)&=&{\cal  A}_1^{(\mp)}+{\cal  A}_2^{(\mp)}(Q^2,\epsilon), 
\end{eqnarray*}
considering $\widetilde{G}_{E,M}^{(\mp)}(Q^2,\epsilon)$ as  {\it generalized} form factors, so that
\begin{equation}
\widetilde{G}_{E,M}^{(\mp)}(Q^2,\epsilon)=\mp G_{E,M}(Q^2)+\Delta G_{E,M}(Q^2,\epsilon).
\label{eq:gem}
\end{equation}
It is important to stress that these eight real functions  completely determine  six complex amplitudes 
${\cal A}_i^{(\mp)}(Q^2,\epsilon)$ 
for $e^{\mp}+N\to e^{\mp}+N$, therefore there are very special connections of 12 real functions $Re{\cal A}_i^{(\mp)}(Q^2,\epsilon)$ and 
$Im{\cal A}_i^{(\mp)}(Q^2,\epsilon)$, with 8 real functions:
\begin{equation}
{\cal A}_i^{(-)}(Q^2,\epsilon)+{\cal A}_i^{(+)}(Q^2,\epsilon)=2 F_i(Q^2),~i=1,2
\label{eq:am1}
\end{equation}
so:
\begin{equation}
Im[{\cal A}_i^{(-)}(Q^2,\epsilon)+{\cal A}_i^{(+)}(Q^2,\epsilon)]=0,~i=1,2
\label{eq:am2}
\end{equation}
in the whole region of space-like momentum transfer.

Note that the C-invariance and the crossing symmetry require that the $\epsilon$-dependence of the six above quoted functions occurs through the argument :
\begin{equation}
x=\sqrt{\displaystyle\frac{1+\epsilon}{1-\epsilon}},
\label{eq:eqx}
\end{equation}
with the following symmetry properties, with respect to the change $x\to -x$:
\begin{eqnarray*}
\Delta G_{E,M}(Q^2,-x)&=& -\Delta G_{E,M}(Q^2,x),\\
{\cal A}_3(Q^2,-x)&=&{\cal A}_3(Q^2,x)
\end{eqnarray*}

These expressions contain the physics of the nucleon electromagnetic structure, taking into account the $1\gamma\bigotimes2\gamma$-interference contribution. In this case, the complete experiment for $eN$-elastic scattering, instead of two real functions $F_{1,2}(Q^2)$, or equivalently, $G_{EN}(Q^2)$, $G_{MN}(Q^2)$, of a single variable $Q^2$, requires six additional functions, depending on two kinematical variables. So, if previously the measurement of the differential cross section, with unpolarized particles in initial and final state, was in principle sufficient, through the Rosenbluth fit, linear in the variable $\epsilon$, now measurements with electrons and positrons in the same kinematical conditions are necessary. The use of the polarization method, which is sensitive to the ratio $G_{EN}/G_{MN}$, allows a better precision at large  momentum transfer.

\section{Analysis of polarization effects}

As in case of the one-photon mechanism, the Breit system is the most convenient  for the general analysis of polarization phenomena for elastic $eN$-scattering. The main difference between the CMS and the Breit system is that, in the last, the three-momenta of the initial and final nucleons are parallel to the three-momentum transfer $\vec q_B$. Therefore we can choose this direction as the $z-$axis and the scattering plane as the $xz$-plane. Neglecting the leptonic mass, $m_e=0$, the components of the four momenta of the initial and scattered electron can be written in the following form:
\begin{eqnarray}
&&~~~~~x~~~,y,~~~z~~~,~~0\nonumber\\
k_1&=\epsilon_B&\left (\cos\displaystyle\frac{\theta_B}{2},0,~\sin\displaystyle\frac{\theta_B}{2},
1~\right )\nonumber\\
k_2&=\epsilon_B&\left (\cos\displaystyle\frac{\theta_B}{2},0,-\sin\displaystyle\frac{\theta_B}{2},
1\right )\nonumber
\label{eq:vkdbe}
\end{eqnarray}
with $k_1^2=k_2^2=0$. The electron energy $\epsilon_B$, where the index $B$ denotes the corresponding variable in the Breit system, can be related with the variables $Q^2$ and $\theta_B$ by:
$$
\epsilon_B=\displaystyle\frac{\sqrt{Q^2}}{2\sin\displaystyle\frac{\theta_B}{2} },~
\cot^2\displaystyle\frac{\theta_B}{2}=\displaystyle\frac
{\cot^2\displaystyle\frac{\theta_e}{2}}{1+\tau}=\displaystyle\frac{2\epsilon}{1-\epsilon}.$$
where $\theta_B$($\theta_e$) is the electron scattering angles in the Breit(LAB) system.

The most direct way to determine the $Q^2$-dependence of the physical FFs 
${G_E}(Q^2)$ and ${G_M}(Q^2)$, in presence of $2\gamma$-contribution, is given by the comparison of $e^-N$ and $e^+N$ scattering. Relations (\ref{eq:gem}) allow to present the differential cross section for $e^{\mp}N$-scattering as follows:
\begin{eqnarray*}
\displaystyle\frac{d\sigma^{(\mp)}}{d\Omega_e}&=&\sigma_0\left \{ \tau\left [G_M^2(Q^2)\mp 2G_M(Q^2) Re \Delta G_M(Q^2)\right ]+\right .\\
&&\hspace*{1true cm}\epsilon\left [G_E^2(Q^2)\mp 2G_E(Q^2) Re \Delta G_E(Q^2)\right ]\\
&&\hspace*{1true cm}\left .\mp 2 \epsilon\sqrt{\tau(1+\tau)\displaystyle\frac{1+\epsilon}{1-\epsilon}}\left [G_E(Q^2)+\tau G_M(Q^2)\right ] Re
{\cal A}_3 \right \},
\end{eqnarray*}
where $(\mp)$ corresponds to $e^{\mp} N$-scattering. One can see that the sum $\Sigma$ of the differential cross sections for $e^{\mp}N$-scattering has the exact Rosenbluth form, in terms of the physical FFs only, i.e. without any $2\gamma$ corrections:
$$\Sigma=\displaystyle\frac{1}{2}\left (\displaystyle\frac{d\sigma^{(-)}}{d\Omega_e}+\displaystyle\frac{d\sigma^{(+)}}{d\Omega_e}\right )=\sigma_0\left [\epsilon G_E^2(Q^2)+\tau G_M^2(Q^2)\right ],$$
which allows to separate $G_E^2(Q^2)$ and $G_M^2(Q^2)$, in presence of the two-photon contribution, without additional assumptions. This can be considered as a generalized Rosenbluth separation.

Therefore, for the physics of electromagnetic FFs, the sum $\Sigma$ of the differential cross sections for $e^{\mp}N$-scattering is relevant, not the difference, as often indicated in the past.

But, as in the case of $1\gamma$-exchange, this generalized Rosenbluth separation is not sufficient to determine the electric FF at high momentum transfer $Q^2$, due to its small relative contribution. Therefore, at large $Q^2$, it is preferable to use polarization observables, such as, for example, the $P_x$ and $P_z$ components of the final nucleon polarization induced by the scattering of longitudinally polarized leptons (electrons and positrons, as well). The corresponding formulas can be written as follows:
\begin{eqnarray*}
{\cal N}P_x^{(\mp)}(Q^2,\epsilon)&=&-\lambda_e\sqrt{2\epsilon(1-\epsilon)\tau}
\left [
Re\widetilde{G}_E^{(\mp)}(Q^2,\epsilon)\widetilde{G}_M^{*(\mp)}(Q^2,\epsilon)
+\right .\\
&&
\left . \sqrt{\tau(1+\tau)}\sqrt{\displaystyle\frac{1+\epsilon}{1-\epsilon}} Re \widetilde{G}_M^{(\mp)}(Q^2,\epsilon)
{\cal A}^*_3(Q^2,\epsilon)\right ]\\
&&\simeq -\lambda_e\sqrt{2\epsilon(1-\epsilon)\tau}\left \{ G_E(Q^2)G_M(Q^2)
\mp\left [ G_E(Q^2) Re \Delta G_M(Q^2,\epsilon)+
 \right .\right .\\
&&\left . \left .G_M(Q^2) Re \Delta G_E(Q^2,\epsilon)\right ]\mp \sqrt{\tau(1+\tau)}\sqrt{\displaystyle\frac{1+\epsilon}{1-\epsilon}}G_M(Q^2) Re {\cal A}_3(Q^2,\epsilon)\right \},\\
{\cal N}P_z^{(\mp)}(Q^2,\epsilon)&=&\lambda_e\tau\sqrt{1-\epsilon^2}
\left [|\widetilde{G}_M^{(\mp)}(Q^2,\epsilon)|^2+
2\epsilon\sqrt{\displaystyle\frac{(1+\tau)\tau}{1-\epsilon^2}}
Re \widetilde{G}_M^{(\mp)}(Q^2,\epsilon){\cal A}^*_3(Q^2,\epsilon)\right ],\\
&&\simeq\lambda_e\tau\sqrt{1-\epsilon^2}\left [G_M^2(Q^2)\mp
2G_M(Q^2)Re \Delta G_M(Q^2,\epsilon)
\right .\\
&&
\left . 
\mp 2\epsilon\sqrt{\displaystyle\frac{(1+\tau)\tau}{1-\epsilon^2}}
G_M(Q^2)Re {\cal A}_3(Q^2,\epsilon)\right ],
\end{eqnarray*}
where $\lambda_e$ is the lepton helicity. Again, the corresponding sum, $\widetilde{P}_x=1/2(P_x^{(-)}+P_x^{(+)})$, $\widetilde{P}_z=1/2(P_z^{(-)}+P_z^{(+)})$, does not contain any contribution from two-photon exchange, and has the known dependence on nucleon form factors \cite{Re68}:
\begin{eqnarray*}
N\widetilde{P}_x&=&-\lambda_e\sqrt{2\epsilon(1-\epsilon)\tau}G_E(Q^2)G_M(Q^2)\\
N\widetilde{P}_z&=&\lambda_e\tau\sqrt{1-\epsilon^2}G_M^2(Q^2).
\end{eqnarray*}
The ratio $\widetilde{P}_x/\widetilde{P}_z$ depends on the ratio of the electric and magnetic nucleon FFs:
$$
\displaystyle\frac{\widetilde{P}_x}{\widetilde{P}_z}=-\sqrt{\displaystyle\frac{2\epsilon}{1+\epsilon}}
\displaystyle\frac{1}{\sqrt{\tau}}\displaystyle\frac{G_E(Q^2)}{G_M(Q^2)}.
$$
Therefore, this method substitutes, in presence of $2\gamma$-exchange, the polarization transfer method \cite{Re68} and it is equally efficient in the extraction of the electromagnetic nucleon FFs.

Actually, beams of electron and positrons, with the same quality are not available at JLab, in spite of the fact that a systematic program of elastic and transition FF measurements is under way. From this point of view, the HERA $e^{\mp}$ beams, where a polarized jet proton target is available, seems well adapted for a deep study of nucleon electromagnetic FFs.

Note, in this respect, that the $E^2$-dependence of the cross section $d\sigma/dQ^2$ (at fixed $Q^2$) results in a factor $[28/(4\div 6)]^2=22\div 49$, increasing of the cross section - in comparison with JLab energies.

\section{T-odd polarization observables}

For the one-photon mechanism, for elastic $eN$-scattering, all T-odd polarization effects vanish in any kinematical condition: beam energy and electron scattering angle. This is a rigorous result which follows from the hermiticity of the nucleon electromagnetic current. One of the consequences is that the nucleon electromagnetic FFs are real functions of the momentum transfer squared in the space-like region. This result holds for positron-nucleon scattering, too.

The situation essentially changes in presence of two-photon exchange. In this case, all three amplitudes, ${\cal A}_i(Q^2,\epsilon)$ [or, equivalently, $\widetilde{G}_E(Q^2,\epsilon)$, $\widetilde{G}_M(Q^2,\epsilon)$, and ${\cal A}_3(Q^2,\epsilon)$] are complex functions of two kinematical variables, and this generates different T-odd polarization phenomena in elastic $eN$-scattering. The simplest, i.e. single spin polarization observables are:

\begin{itemize}
\item $P_y$: the transversal component of the final nucleon polarization, in the collision of unpolarized leptons with unpolarized nucleon target, $e^{\mp}+N\to 
e^{\mp}+\vec N$;
\item $A_y$: the analyzing power, induced by the collision of unpolarized leptons (electrons or positrons) with a transversally polarized nucleon target,
$e^{\mp}+\vec N\to e^{\mp}+ N$;

\item ${\cal A}^{(e)}$: the asymmetry in the scattering of a transversally polarized lepton beam  with an unpolarized nucleon target, $\vec e^{\mp}+N\to e^{\mp}+ N$.
\end{itemize}
Note, in this respect, that polarization phenomena in $eN$-interaction, induced by a transverse lepton polarization, are generally smaller, by a factor $\simeq m_e/E_e$ ($m_e$ and $E_e$ are the mass of the electron and its energy), in comparison with the polarization effects induced by the longitudinal lepton polarization. Therefore the asymmetry $A^{(\mu)}$ for muon-nucleon scattering is essentially larger ($m _{\mu}/m_e\simeq 200$). However, typically, the muons have only longitudinal polarization, which is induced by the helicity conservation in the decays $\pi^+\to\mu^+ +\nu_{\mu}$ or $\pi^-\to\mu^- +\overline{\nu}_{\mu}$.

The P-invariance of the electromagnetic interaction of nucleons results in the fact that double-spin polarization observables in elastic $eN$-scattering, such as the components $P_x$ and  $P_z$ of the final nucleon polarization (in collisions of a longitudinally polarized lepton beam with unpolarized nucleon target), $\vec e^{\mp}+N\to e^{\mp}+ \vec N$ or the asymmetries $A_x$ and $A_z$ (characterizing the collisions of a longitudinally polarized lepton beam with a polarized nucleon target, in the reaction plane, $\vec e^{\mp}+\vec N\to e^{\mp}+ N$) must be T-even. For the same reasons, the following components of the depolarization tensor, $D_{ab}$, $a,b=x,y,z$, which characterize the  dependence of the $b$-component of the final nucleon polarization on the $a$-
component of the  nucleon target, (in the scattering of unpolarized leptons  
$e^{\mp}+\vec N\to e^{\mp}+ \vec N$):
$$D_{xx},~D_{yy},~D_{zz},~D_{xz},~\mbox{and~}D_{zx},$$
are T-even. Therefore, these observables do not vanish in case of one-photon mechanism. But the components, $D_{xy}$, $D_{yx}$, $~D_{zy}$, and $D_{yz}$ are identically zero, even in presence of two-photon exchange, due to the P-invariance of the electromagnetic hadronic interaction. This means that all double-spin polarization phenomena in elastic $eN$-scattering are of T-even nature.

So, only triple-spin polarization observables in elastic $eN$-scattering, such as the components of the depolarization tensor, induced by the scattering of longitudinally polarized leptons:
$$D_{xy}(\lambda_e),~D_{yx}(\lambda_e),~D_{zy}(\lambda_e),~\mbox{and~}D_{yz}(\lambda_e),$$
are T-odd. 

All these T-odd polarization observables are generally determined by the following three independent combinations of amplitudes:
\begin{eqnarray}
{\cal I}_1(Q^2,\epsilon)&= &Im \widetilde{G}_E(Q^2,\epsilon){\cal A}_3^*(Q^2,\epsilon)\simeq\mp {G_E}(Q^2)Im{\cal A}_3^*(Q^2,\epsilon), \nonumber\\
{\cal I}_2(Q^2,\epsilon)&= &Im \widetilde{G}_M(Q^2,\epsilon){\cal A}_3^*(Q^2,\epsilon)\simeq\mp {G_M}(Q^2)Im{\cal A}_3^*(Q^2,\epsilon), \label{eq:eqi13}\\
{\cal I}_3(Q^2,\epsilon)&= &Im \widetilde{G}_E(Q^2,\epsilon) \widetilde{G}_M^*(Q^2,\epsilon)\simeq\mp [{G_E}(Q^2)Im\widetilde{G}_M^*(Q^2,\epsilon)
+{G_M}(Q^2)Im\widetilde{G}_E(Q^2,\epsilon)]. \nonumber
\end{eqnarray}

It follows that all T-odd polarization observables are equal in absolute value and differ in sign for $e^-N$ and  $e^+N$ scattering,
in framework of $1\gamma\bigotimes 2\gamma$ mechanism, keeping the $\alpha^2$ (main) and $\alpha^3$ (correction) contributions.
Moreover, and this is the main result of this analysis, the ratio 
$${\cal R}={\cal I}_1(Q^2,\epsilon)/{\cal I}_1(Q^2,\epsilon)={G_E}(Q^2)/{G_M}(Q^2),$$
being $\epsilon$-independent, (at any value of $Q^2$ and initial electron energy) allows to determine the necessary ratio of the electric and magnetic form factors, independently of any assumption, concerning $2\gamma$-exchange.

Note that the ratio ${\cal R}(Q^2)$, being function of the momentum transfer squared, only, has to be the same for electron and positron scattering, for the $1\gamma\bigotimes 2\gamma$ mechanism.

Therefore, the measurement of the necessary set of T-odd polarization phenomena can be considered as the method for the determination of the ratio of the real, physical, nucleon form factors, ${G_E}(Q^2)/{G_M}(Q^2)$, not of the generalized form factors, unknown functions of $Q^2$ and $\epsilon$. In this way, the physics of electromagnetic nucleon FFS is preserved, although it requires more difficult experiments of percent precision, as the T-odd polarization effects are of the order of $\alpha$. However, from a theoretical point of view, it is a transparent method, free from  a priori unjustified approximations.

Let us derive the explicit expressions for the above mentioned T-odd polarization effects, in terms of 
${\cal I}_1(Q^2,\epsilon)$, defined in (\ref{eq:eqi13}):
\begin{eqnarray}
{\cal N}P_y={\cal N}A_y&=&-\sqrt{2\epsilon\tau(1+\epsilon)} \left[ {\cal I}_3(Q^2,\epsilon)-
\displaystyle\frac{2\epsilon}{1+\epsilon} {\cal I}'_1(Q^2,\epsilon)+{\cal I}'_2(Q^2,\epsilon)\right ], \label{eq:eq1}\\
{\cal N}D_{xy}(\lambda_e)={\cal N}D_{yx}(\lambda_e)&=&-2\lambda_e\epsilon\tau
\sqrt{\displaystyle\frac{1-\epsilon}{1+\epsilon}}{\cal I}'_2(Q^2,\epsilon),\label{eq:eq2}\\
{\cal N}D_{yz}(\lambda_e)=-{\cal N}D_{zy}(\lambda_e)
&=&-\lambda_e\sqrt{2\epsilon\tau(1-\epsilon)}\left[ {\cal I}_2'(Q^2,\epsilon)+{\cal I}_3(Q^2,\epsilon)\right ],\label{eq:eq3}\\
{\cal N}{\cal A}^{(e)}&=&S_y^{(e)}\sqrt{2\epsilon(1+\tau)(1-\epsilon)}~\displaystyle\frac{m_e}{m}\tau~{\cal I}_2(Q^2,\epsilon),\label{eq:eq3a}
\end{eqnarray}
where 
$${\cal I}'_{1,2}(Q^2,\epsilon)={\cal I}_{1,2}(Q^2,\epsilon)\sqrt{\tau (1+\tau) \displaystyle\frac{1-\epsilon}{1+\epsilon}},$$
$${\cal N}=|\widetilde{G}_E(Q^2,\epsilon)|^2+\tau |\widetilde{G}_M(Q^2,\epsilon)|^2
+2\epsilon\sqrt{\tau (1+\tau)\displaystyle\frac{1+\epsilon}{1-\epsilon}}~ Re [ G_E(Q^2)+\tau G_M(Q^2)]{\cal A}_3(Q^2,\epsilon),
$$
and $S_y^{(e)}$ is the transversal component of the lepton polarization.

From these formulas, can see that ${\cal A}^{(e)}\simeq m_e/m\simeq 10^{-3}$, as it results from the transversal component of the lepton spin. Moreover the following relation holds:
$${\cal A}^{(e)}=S_y^{(e)}\displaystyle\frac{m_e}{M}\tau\sqrt{\displaystyle\frac{2}{\epsilon}(1-\epsilon)}\left [ \displaystyle\frac{D_{xy}(\lambda_e)}{\lambda_e}\right ].$$
This is a model independent result, which holds in any kinematical condition.

For a model independent determination of of the nucleon electromagnetic form factors, $G_E(Q^2)$ and $ G_M(Q^2)$, it is necessary to measure not only the simplest (single-spin) polarization observables $P_y$ and $A_y$, but also the triple-spin polarization observables
$D_{zy}(\lambda_e)$ and $D_{yz}(\lambda_e)$, in order to determine the three functions $){\cal I}_i(Q^2,\epsilon)$.
The corresponding solution can be written as:
\begin{eqnarray*}
{\cal I}_1'(Q^2,\epsilon)&=&\displaystyle\frac{1}{2\lambda_e}\sqrt{\displaystyle\frac{1+\epsilon}{\epsilon\tau}}
\left [ P_y+ \sqrt{\displaystyle\frac{1+\epsilon}{1-\epsilon}}
\displaystyle\frac{D_{zy}(\lambda_e)}{\lambda_e}\right ],\\
{\cal I}'_2(Q^2,\epsilon)&=&\displaystyle\frac{1}{2\epsilon\tau}
\sqrt{\displaystyle\frac{1+\epsilon}{1-\epsilon}}\displaystyle\frac{D_{xy}(\lambda_e)}{\lambda_e},\\
{\cal I}'_3(Q^2,\epsilon)&=&\displaystyle\frac{1}{\sqrt{2\epsilon\tau(1-\epsilon)}}
\left [\displaystyle\frac{D_{zy}(\lambda_e)}{\lambda_e}+ \displaystyle\frac{1}{\sqrt{2\epsilon\tau}}
\displaystyle\frac{D_{xy}(\lambda_e)}{2\lambda_e}\right ].\\
\end{eqnarray*}

Summarizing the discussion in this section, devoted to the analysis of the T-odd polarization effects in elastic $eN$-scattering, we showed that the measurement of the ratio 
\begin{equation}
\displaystyle\frac{G_E(Q^2)}{G_M(Q^2)}=
\displaystyle\frac{{\cal I}_1(Q^2,\epsilon)}{{\cal I}_2(Q^2,\epsilon)}=-
\displaystyle\frac{\lambda_e}{D_{xy}(\lambda_e)}
\left [ P_y+\sqrt{\displaystyle\frac{1+\epsilon}{1-\epsilon}}
\displaystyle\frac{D_{zy}(\lambda_e)}{\lambda_e}\right ]
\sqrt{\tau\displaystyle\frac{1-\epsilon}{2\epsilon}}
\label{eq:eq11}
\end{equation} 
involves three T-odd observables, one single-spin observable, $P_y$ or $A_y$, and two triple-spin polarization observable $D_{yz}(\lambda_e)$, and $D_{zy}(\lambda_e)$. So, the combination (\ref{eq:eq11}) represents another possible way for the determination of the nucleon electromagnetic FFs, in presence of the $2\gamma$ contribution. It is, also, a model independent way.

\section{T-even effects in $e^-p$--scattering}
For completeness we discuss here the possibility to extract nucleon electromagnetic FFs, from polarization observables in elastic $e^-p$-scattering, in absence of a positron beam. In principle, this problem can be solved, by measuring T-even polarization observables, such as the differential cross section, the $P_{x,z}$ components of the final nucleon polarization, the components of the depolarization tensor etc.

In the general case, all these observables are determined by the following quadratic combinations of the amplitudes $\widetilde{G}_{E,M}(Q^2,\epsilon)$ and ${\cal A}_3(Q^2,\epsilon),$ (taking into account contributions in $\alpha^2$ and $\alpha^3$):
$$
|\widetilde{G}_E^(Q^2,\epsilon)|^2,~|\widetilde{G}_M(Q^2,\epsilon)|^2,~
Re \widetilde{G}_E(Q^2,\epsilon) \widetilde{G}_M^*(Q^2,\epsilon),
$$
\begin{equation}
R_1=G_E(Q^2)Re {\cal A}_3(Q^2,\epsilon),~\mbox{and~}R_2=G_M(Q^2)Re {\cal A}_3(Q^2,\epsilon).
\label{eq:eq12}
\end{equation}
The most interesting quantities are related to $R_{1,2}$, because the ratio
$$\displaystyle\frac{R_1}{R_2}=\displaystyle\frac{G_E(Q^2)}{G_M(Q^2)}$$
allows to determine, in another independent way, the ratio of electromagnetic form factors in presence of $2\gamma$-exchange.

The measurement of the differential cross section, $P_x$ and $P_z$ is not sufficient for the determination of the ratio $R_1/R_2$, because these observables depend on five combinations of amplitudes $\widetilde{G}_{E,M}(Q^2,\epsilon)$ and ${\cal A}_3(Q^2,\epsilon)$, Eq. (\ref{eq:eq12}). To avoid additional assumptions, which may be difficult to justify, it is necessary to measure two additional polarization observables, of T-even nature. Note that two possible double-spin observables, $A_x$ and $A_z$ (describing the asymmetry of longitudinally polarized electron scattering on a polarized proton target, in $x$ and $z$ direction, respectively) do not help, because they do not contain additional information with respect to $P_x$ and $P_z$, as the following relations hold:
$$A_x=P_x,~A_z=-P_z,$$
in presence of $1\gamma$ and $2\gamma$ exchange, as well.

Therefore the components of the depolarization tensor $D_{ab}$ have also to be considered. For the scattering of unpolarized electrons, taking into account the P-invariance in the electromagnetic interaction of nucleons, the following components generally do not vanish:
$$D_{xx},~D_{yy},~D_{zz},~D_{xz},~\mbox{and}~D_{zx}.$$
We can decompose  each of these components in two parts:
$$D_{ab}=D_{ab}^{(0)}+D_{ab}^{(1)},$$
where $D_{ab}^{(0)}$ describes the main contribution in terms of $|\widetilde{G}_{E,M}|^2$ and 
$Re\widetilde{G}_E\widetilde{G}_M^*$, which is equivalent to the expression in frame of $1\gamma$-exchange:
\begin{eqnarray}
{\cal N}D_{xx}^{(0)}&=&\epsilon\left [ |\widetilde{G}_E(Q^2,\epsilon)|^2-\tau|\widetilde{G}_M(Q^2,\epsilon)|^2\right ],\nonumber\\
{\cal N}D_{yy}^{(0)}&=&\epsilon\left [ |\widetilde{G}_E(Q^2,\epsilon)|^2+\tau|\widetilde{G}_M(Q^2,\epsilon)|^2\right ],\nonumber\\
{\cal N}D_{zz}^{(0)}&=&\epsilon|\widetilde{G}_E(Q^2,\epsilon)|^2-\tau|\widetilde{G}_M(Q^2,\epsilon)|^2,\nonumber\\
{\cal N}D_{zx}^{(0)}&=&-{\cal N}D_{xz}^{(0)}=\sqrt{2\epsilon\tau(1+\epsilon)}Re \widetilde{G}_E(Q^2,\epsilon)
\widetilde{G}_M^*(Q^2,\epsilon)\label{eq:eq13}
\end{eqnarray}
and the part depending on $2\gamma$-exchange:
\begin{eqnarray}
{\cal N}D_{xx}^{(1)}&=&{\cal N}D_{zz}^{(1)}=\sqrt{1+\epsilon}\left [ R_1'(Q^2,\epsilon)-\tau R_2'(Q^2,\epsilon)\right ],\nonumber\\
{\cal N}D_{yy}^{(1)}&=&\sqrt{1+\epsilon}\left [ R_1'(Q^2,\epsilon)+\tau R_2'(Q^2,\epsilon)\right ],\nonumber\\
{\cal N}D_{zx}^{(1)}&=&-{\cal N}D_{xz}^{(1)}=\sqrt{2\epsilon}\left [ R_1'(Q^2,\epsilon)+
\displaystyle\frac{1+\epsilon}{2\epsilon} R_2'(Q^2,\epsilon)\right ],\label{eq:eq14}
\end{eqnarray}
where
$$
R'_{1,2}(Q^2,\epsilon)=\displaystyle\frac{2\epsilon}{\sqrt{1-\epsilon}}\sqrt{\tau(1+\tau)}
G_{E,M}(Q^2) Re{\cal A}_3(Q^2,\epsilon).
$$

The relation $D_{xz}=-D_{zx}$, which is valid for $1\gamma$ exchange, still holds, after including the $2\gamma$ contribution, as well as the following relation among the diagonal elements of the depolarization tensor $D_{ab}$:
$$ D_{xx}+D_{yy}-D_{zz}=1.$$

These formulas show that at least five different measurements must be performed, in order to determine the necessary combinations $R'_{1,2}(Q^2,\epsilon)$. For example, the measurements of 
$d\sigma/d{\Omega}$, $P_x$, $P_z$, $D_{xz}$ and $D_{xx}$ will result in a system of five linear equations, with respect to five functions, see Eq. (\ref{eq:eq12}).

The unique solution of this system, at any $Q^2$ and $\epsilon$ allows to determine all quantities (\ref{eq:eq12}). The important physical information about nucleon electromagnetic FFs is contained in $R_1'$ and $R_2'$ from which it is possible to extract the ratio $G_E(Q^2)/G_M(Q^2)$:
\begin{equation}
\displaystyle\frac{G_E(Q^2)}{G_M(Q^2)}=
\displaystyle\frac{1}{2\epsilon}
\sqrt{\tau\displaystyle\frac{(1-\epsilon^2)}{2\epsilon}}
\left (P_x\sqrt{1+\epsilon}-D_{xz}\sqrt{1-\epsilon}\right )
\left [-1-\displaystyle\frac{1-\epsilon}{2\epsilon}D_{xx}
+ \displaystyle\frac{1+\epsilon}{2\epsilon}D_{yy}\right ]^{-1}.
\label{eq:eq15}
\end{equation}
which involves five different T-even observables. Evidently, all these observables do not vanish in $1\gamma$ exchange, but the main contributions cancel in the ratio, Eq. (\ref{eq:eq15}). Therefore, Eq. (\ref{eq:eq15})
constitues the third way for the measurement of the nucleon FFs, but, again, one has to deal with very small contributions.

Finally, this indicates that the $\epsilon$ dependence of all these observables is not so important, for this aim.  Note, however, that  $R_1'$ and $R_2'$ are small in absolute value, being proportional to $\alpha$, and the measurement of the corresponding polarization observables requires adequate accuracy.
The determination of the "generalized" form factors, is not of interest, because these functions can not be easily interpreted in terms of nucleon structure. Moreover, in the general case, these FFs depend strongly on the parametrization of the spin structure for the third (i.e. non vector) amplitude in the general matrix element, Eq. (\ref{eq:mat1}).

\section{$2\gamma$-effects in other processes}

We proved above that the $2\gamma$ exchange essentially modify the analysis of polarization phenomena in elastic lepton-nucleon scattering, so that the extraction of physical form factors $G_{E,M}(Q^2)$ has a larger complexity, in comparison with the $1\gamma$ mechanism.

Let us briefly discuss here other processes of elastic and inelastic scattering of leptons (positron or electrons) bt hadrons.

The simplest case is  the elastic $e^{\pm}$ scattering by hadronic target with zero spin, as for example,  $e^{\pm}+^4\!He\to e^{\pm}+^4\!He$-scattering. The following analysis is also valid for inverse kinematics, i.e. for scattering of high energy $\pi$ or $K$ mesons by atomic electrons, $\pi(K)+e^-\to\pi(K)+e^- $.

Taking into account the electron helicity conservation, the matrix element for 
$e^{\pm}+h\to e^{\pm}+h$, $h$ is a spin zero hadron, can be written in the following general form:
$${\cal M}=2\displaystyle\frac{e^2}{Q^2}{\cal A}^{(\pm)}(Q^2,\epsilon)\overline u(k_2)\hat{\cal P}u(k_1),$$
where ${\cal A}^{(\pm)}(Q^2,\epsilon)$ is the (single) invariant amplitude  for the considered process. Considering $1\gamma\bigotimes 2\gamma$-exchanges, one can write: 
$$
{\cal  A}^{(\mp)}(Q^2,\epsilon)= \mp {\cal F}(Q^2)+\Delta {\cal  A}(Q^2,\epsilon),
$$
where ${\cal F}(Q^2)$ is the corresponding form factor for a spinless hadron $h$, which is a real function of $Q^2$ (in the space-like region).

As a result of C-invariance and the crossing symmetry, $\Delta {\cal  A}(Q^2,\epsilon)$, depends on  $\epsilon$ through the same argument $x$, Eq. (\ref{eq:eqx}), with the following symmetry relation: $\Delta {\cal  A}(Q^2,-x)=-\Delta {\cal  A}(Q^2,x)$.

The presence of a single amplitude ${\cal  A}^{(\mp)}(Q^2,\epsilon)$ in the matrix elements, results in definite values for the polarization phenomena, which is not interesting for the extraction of form factors. Therefore, the method to disentangle the reaction mechanism, in this case, is only to measure the sum of the differential cross section for the scattering of positrons and electrons, in the same kinematical conditions.

The situation is essentially more complicated with a spin one target, as, for example, $ed$-scattering. The conservation of helicity, in elastic $ed$-scattering, generates the following independent helicities transitions (taking into account the identity of the initial and final states):
\begin{equation}
\begin{array}{lll}
e~+~d  &~\to~ e~+&d\nonumber\\
+~~~~ + &~\to~ + &+\nonumber\\
+~~~~ + &~\to~ + &0\nonumber\\
+~~~~ + &~\to~ + &-\nonumber\\
+~~~~ 0 &~\to~ + &0\nonumber\\
+~~~~ 0 &~\to~ + &-\nonumber\\
+~~~~ - &~\to~ + &-\nonumber\\
\end{array}
\end{equation}
where $+$ indicates, for the electron, $\lambda_e=+1/2$, and $+,0$ and $-$, indicate, for the deuteron, helicity
$\lambda_d=+1$, $0$ and $-1$, respectively. Therefore, for $ed$-elastic scattering there are six independent complex amplitudes,
$A_{i,d}^{(\pm)}$, $i=1-6$, for electron and positron scattering, i.e. four times more in comparison with one-photon exchange (with three real electromagnetic form factors, functions of $Q^2$, only). More precisely, the C-invariance requires that both processes 
$e^{\mp}+d\to e^{\mp}+d $ are described by a set of nine real functions, three of which can be identified with the deuteron form factors, and six functions, depending on $\epsilon$ and $Q^2$, which contain the $2\gamma$-contribution.

In principle it is possible to analyze  $ed$ elastic scattering, as it has been done above for $eN$ scattering, and to indicate which  polarization observables are necessary to extract the electromagnetic form factors. As an example, to measure all the independent T-odd combinations, it is necessary to measure at least twelve polarization observables, with large accuracy. As in case of $eN$ elastic scattering, the necessary nine combinations are of the order of $\alpha$. This reaction, with spin one particles in initial and final states, offers a large number of polarization observables. Up to now, three polarization experiments have been realized for $ed$ scattering:
the measurement of the deuteron tensor polarization, the measurement of the asymmetry, with a tensor polarized target, and the measurement of a T-odd vector polarization.

The situation with T-even polarization observables is even more complicated, with fifteen necessary polarization experiments.

Therefore, the positron method seems, again, as the most realistic: in addition to the sum of the differential cross sections, with the Rosenbluth separation of the structure functions $A(Q^2)$ and $B(Q^2)$, it is necessary to measure also the tensor deuteron polarization in elastic $e^+d$ and $e^-d$ collisions, in order to determine the contributions of the electric and quadrupole deuteron form factors.

For the processes $e^{\pm}+N\to e^{\pm}+N^*$ (with excitation of nucleonic resonances, with different $J^P$), the presence of a two-photon contribution makes the extraction of the transition FFs even more complicated, due to the difference in the initial and final hadronic systems, resulting in a very large number of additional two-photon structures. It is possible to find that, for any process 
 $e^{\pm}+N\to e^{\pm}+N^*(J)$ one has 
 $$2\times (2J+1)=4J+2$$
 complex amplitudes (functions of $ Q^2$ and $\epsilon$). For example, for $e^{\pm}+N\to e^{\pm}+\Delta^+(J^P=3/2^+)$, five additional complex amplitudes, due to $ 2\gamma$ exchange are present, increasing the complexity of the corresponding analysis of polarization phenomena.
 
Note, in this respect, that, the presence of the two-photon contribution would affect particularly the study of  $e^{\pm}+N\to e^{\pm}+\Delta(1232)$, as two  FFs for the vertex $\gamma^++N\to \Delta$ (the longitudinal and transversal elastic quadrupole ones) are very small. These FFs are very important for testing pQCD, with respect, for example to the understanding of hadron helicity conservation \cite{Le80}. The last data \cite{Fr99} are in contradiction with the pQCD prediction.

Concluding this section, we stress that the positron method seem the most promising to extract information on inelastic electromagnetic FFs, too. The presence of the two-photon contribution prevents the description of the processes $e^{\pm}+N\to e^{\pm}++N^*(J^P)$, in terms of three FFs only, and requires a careful analysis of each process, with a complicated spin structure of the corresponding matrix element.

\section{Conclusions}
The general symmetry properties of electromagnetic interaction, such as the C-invariance, the crossing symmetry and the lepton helicity conservation in QED, allow to obtain  rigorous results concerning two-photon exchange contributions for elastic $eN$-scattering and to analyze the effects of this mechanism in $eN$-phenomenology \cite{Re03}.

The form factors $G_{EN}(Q^2)$ and $G_{MN}(Q^2)$ and the $2\gamma$-amplitudes, $\Delta G_{E,M}(Q^2,\epsilon)$ are the same for $e^+p$ and $e^-p$ elastic scattering. This allows to connect the difference of the differential cross sections for $e^{\mp} p$-interaction with the deviations from the $\epsilon$-linearity of the Rosenbluth plot.

The $\epsilon$-dependence of the interference contribution to the differential cross section of $e^{\mp}p$ elastic scattering is very particular. Any approximation of this term by a linear function in the variable $\epsilon$ is in contradiction with C-invariance and crossing symmetry of the electromagnetic interaction. 

We analyzed above different possible strategies in the determination of the nucleon electromagnetic FFs, $G_E(Q^2)$ and $G_M(Q^2)$, through the measurement of different polarization observables in elastic $e^{\pm} N$ scattering, of T-even and T-odd nature, in presence of $2\gamma$ exchange.

There are in principle three different ways , to determine the physical nucleon FFs, $G_{E,M}(Q^2)$. The proposed methods are, all, relatively complicated but allow to go beyond the description in terms of 'generalized' FFs, which are functions of two kinematical variables, $Q^2$ and $\epsilon$ and are not directly related to the nucleon electromagnetic structure.

The formally simplest way needs the parallel study of positron and electron scattering, in the same kinematical conditions. We showed that the two-photon contribution cancels in the sum of the differential cross sections, $d\sigma^{(-)}/d\Omega_e+d\sigma^{(+)}/d\Omega_e$. A linear $\epsilon$-fit of this quantity allows to extract $G_E(Q^2)$ and $G_M(Q^2)$, through a generalized Rosenbluth separation. At higher $Q^2$, due to the small contribution of $G_E(Q^2)$,  the polarization transfer method should be used, which requires the measurement of the $P_x$ and $P_z$-components of the final nucleon polarization -with longitudinally polarized electron and positron beams. This can be in principle realized at the HERA $e^{\pm}$ ring, with a polarized jet proton target.

In absence of positron beam two other possibilities to measure $G_{E,M}(Q^2)$ can be suggested, using only an electron beam.

One possibility is the measurement of T-odd polarization observables, such as $P_y$, $D_{zy}(\lambda_e)$, and $D_{yz}(\lambda_e)$ (i.e. the components of the depolarization tensor, in the scattering of longitudinally polarized electrons by a polarized target, with the measurement of the final proton polarization). All these observables, which vanish in the Born approximation for $eN$-scattering, must be of  the order of $\alpha$ and they should be measured with corresponding accuracy.

Another possible way requires the measurement of T-even polarization observables. The general spin structure of elastic $eN$-scattering, made more complicated by the presence of the $2\gamma$-contributions, contains three independent complex amplitudes (instead of two real terms, in case of $1\gamma$ approximation). This requires the determination of five quadratic combinations of these amplitudes. Only two of them, $G_E(Q^2)Re {\cal A}_3(Q^2,\epsilon)$ and $R_2=G_M(Q^2)Re {\cal A}_3(Q^2,\epsilon)$ are relevant for the extraction of the FFs, and can be found through the measurement of five T-even observables, such as $d\sigma/d\Omega_e$, $P_x$ (or $A_x$), and the $D_{xx}$, $D_{yy}$ and $D_{xz}$ components of the depolarization tensor (for unpolarized electron scattering).

Therefore, in presence of a $2\gamma$-contribution, the extraction of the nucleon electromagnetic FFs is still possible, but requires more complicated experiments, with a very high level of precision. Only in this way it will be possible to investigate the nucleon structure, at large momentum transfer, keeping the elegant formalism of QED, traditionally used for this aim.

{}

\begin{thebibliography}{}

\bibitem{Ho62} 
R.~Hofstadter, F Bumiller and M. Yearian,
 Rev.\ Mod. Phys. \ {\bf 30}, 482 (1958).

\bibitem{Shwinger}
J.~S.~Schwinger,
Phys.\ Rev.\  {\bf 76},  790(1949).



\bibitem{Gu73}  J. Gunion and L. Stodolsky,  Phys. Rev. Lett. {\bf 30}, 345
(1973);\\
 V. Franco,  Phys. Rev. D {\bf 8}, 826 (1973); \\
 V. N. Boitsov, L.A. Kondratyuk and V.B. Kopeliovich, Sov. J.
Nucl. Phys. {\bf 16}, 237 (1973); \\
  F. M. Lev, Sov. J. Nucl. Phys. {\bf 21}, 45 (1973).

\bibitem{Re99} M. P. Rekalo, E. Tomasi-Gustafsson and D. Prout, Phys. Rev.
{\bf  C60}, 042202 (1999).

\bibitem{Al99} L. C. Alexa {\it et al.}, Phys. Rev.  Lett. {\bf 82}, 1374 
(1999).
\bibitem{Ab99} D. Abbott {\it et al.}, Phys. Rev. Lett {\bf 82}, 1379 (1999).

\bibitem{Re03} M. P. Rekalo and E. Tomasi-Gustafsson arXiv:hep-ph/0310172.

\bibitem{Jo00}
M.~K.~Jones {\it et al.}  [Jefferson Lab Hall A Collaboration],
Phys.\ Rev.\ Lett.\  {\bf 84}, 1398 (2000);

O.~Gayou {\it et al.}  [Jefferson Lab Hall A Collaboration],
Phys.\ Rev.\ Lett.\  {\bf 88}, 092301 (2002).
\bibitem{Re68} A. Akhiezer and M. P. Rekalo, Dokl. Akad. Nauk USSR, {\bf 180},
1081 (1968); Sov. J. Part. Nucl. {\bf 4}, 277 (1974).
\bibitem{Ar75} R.G. Arnold {\it et al.}, Phys. Rev. Lett. {\bf 35}, 776 (1975).
\bibitem{Br03} S.J. Brodsky  {\it et al.}, arXiv:hep-ph/0310277 and refs. herein.
\bibitem{Mo69} L. W. Mo and Y. S. Tsai, Rev. Mod. Phys. {\bf 41}, 205 (1969).
\bibitem{Ma00} 
L.~C.~Maximon and J.~A.~Tjon,
Phys.\ Rev.\ C {\bf 62}, 054320 (2000);\\
A.~Afanasev, I.~Akushevich and N.~Merenkov,
Phys.\ Rev.\ D {\bf 64}, 113009 (2001);\\
A.~V.~Afanasev, I.~Akushevich, A.~Ilyichev and N.~P.~Merenkov,
Phys.\ Lett.\ B {\bf 514}, 269 (2001).
\bibitem{Go57} M. L. Goldberger, Y. Nambu and R. Oehme, Annals of Physics {\bf 2}, 226 (1957).
\bibitem{Le80} G. P. Lepage and S. J. Brodsky, Phys. Rev. D {\bf 22}, 2157 (1980).
\bibitem{Fr99} V. V. Frolov  {\it et al.}, Phys. Rev. Lett. {\bf 82}, 45 (1999).
\end{thebibliography}
\end{document}